\documentclass[aps,pra,showpacs,preprint]{revtex4}

\usepackage{amssymb, color}
\usepackage{amsmath}
\usepackage{amsfonts}
\usepackage{epsfig,subfigure,dsfont,amsthm,amsbsy,mathrsfs,amscd}

\usepackage{enumerate}
\usepackage{graphicx}
\usepackage{url}
\usepackage{bm}
\usepackage{multirow}
\usepackage{pifont}

\usepackage{float}

\def\la{\langle}
\def\ra{\rangle}
\def\be{\begin{equation}}
\def\ee{\end{equation}}
\def\ba{\begin{array}}
\def\ea{\end{array}}

\newcommand\btd{\raise 2pt \hbox{$\hat\bigtriangledown$}\hskip 1.5pt}
\newcommand\bt{\raise 2pt \hbox{$\bigtriangledown$}\hskip 1.5pt}
\newcommand{\ot}{\otimes}

\begin{document}

\title{Time optimal control based on classification of quantum gates}

\author{Bao-Zhi Sun}
\email{sunbaozhi@qfnu.edu.cn}
\affiliation{School of Mathematical Sciences, Qufu Normal University, Shandong 273165, China}

\author{Shao-Ming Fei}
\email{feishm@cnu.edu.cn}
\affiliation{School of Mathematical Sciences,  Capital Normal University,  Beijing 100048,  China}
\affiliation{Max-Planck-Institute for Mathematics in the Sciences, Leipzig 04103, Germany}

\author{Naihuan Jing}
\email{jing@ncsu.edu}
\affiliation{Department of Mathematics, North Carolina State University, Raleigh, NC, USA}

\author{Xianqing Li-Jost}
\email{xianqing.li-jost@mis.mpg.de}
\affiliation{Max Planck Institute for Mathematics in the Sciences, Leipzig, Germany}

\begin{abstract}
We study the minimum time to implement an arbitrary two-qubit gate in two heteronuclear spins systems.
We give a systematic characterization of two-qubit gates based on the invariants of local equivalence.
The quantum gates are classified into four classes, and for each class the analytical formula of the
minimum time to implement the quantum gates is explicitly presented.
For given quantum gates, by calculating the corresponding invariants one easily obtains the classes to which the quantum gates belong.
In particular, we analyze the effect of global phases on the minimum time to implement the gate.
Our results present complete solutions to the optimal time problem in implementing an arbitrary two-qubit gate in two heteronuclear spins systems.
Detailed examples are given to typical two-qubit gates with or without global phases.
\end{abstract}


\maketitle

\section{Introduction}
Optimal control of a quantum system \cite{Warren,Rabitz,Daniel,Navin2001,ZhJun2003,Yu2013,TOC-2013-Garon} plays an important role in quantum computation and quantum information processing \cite{Nielsen},
as any physical design of a quantum computer must be able to realize a set of quantum gates for computational purpose. Then it is a practical problem to know how quickly the quantum system can carry out such tasks
both heuristically and theoretically.
However, it has been a challenging problem to determine the minimum time analytically for implementing an arbitrary unitary transformation.
Based on the Cartan decomposition of the unitary operator, the authors in \cite{Navin2001} presented an elegant analytical characterization of the minimum time required to steer
the system from an initial state to a specified final state for a given controllable right invariant system, described by certain Hamiltonian with both a local control and nonlocal internal or drift terms.
However, since the Cartan decompositions of a unitary operator are not unique, operationally it is quite difficult to compute the minimal time for a given quantum gate.

In \cite{ZhJun2003} local invariants were introduced for the equivalence of unitary operators under local transformations.
Based on these local invariants, an operational approach \cite{Yu2013} was given to compute the minimal time required to implement a given unitary operator for the heteronuclear system
\cite{Glaser1998}. 
Unfortunately, the results given in \cite{Yu2013} are not completely correct, and can even not distinguish the minimal time required to implement a gate
and that required to implement the same gate with a global phase.

A state $\rho(0)$ of a quantum system at time zero evolves into the state $\rho(t)$ at time $t$ in such a way that $\rho(t)=U(t)\rho(0) U^\dag(t)$ for some unitary operator $U(t)$, where
the unitary operator $U(t)$ is determined by the Hamiltonian $H(t)$ of the system, satisfying the time-dependent Schr\"{o}dinger equation,
$\dot{U}(t)=-iH(t)U(t)$,
with $U(0)=I$ the identity operator.


For systems of two heteronuclear spins coupled by a scalar $J$ \cite{Glaser1998}, assuming each spin can be excited individually,
the control problem is to implement any given unitary transformation $U\in \mathrm{SU}(4)$ from the specified coupling and single-spin operations, the case appears often in the nuclear magnetic resonance (NMR) systems.
The unitary propagator $U$ is governed by the following equation,
\be\label{Heter-Uni}
\dot{U}(t)=-i(H_d+\sum_{k=1}^4v_j(t)H_j)U(t), \qquad U(0)=I,
\ee
where $H_d=\frac{\pi}{2} J \sigma_z\otimes \sigma_z$, $H_1=\pi\sigma_x\otimes I$,
$H_2=\pi\sigma_y\otimes I$, $H_3=\pi I\otimes\sigma_x$, $H_4=\pi I\otimes\sigma_y$, with $\sigma_x$, $\sigma_y$ and $\sigma_z$ the Pauli matrices, and $I$ the identity operator.
$J$ is the coupling strength between the spins.
All the unitary gates belonging to $\mathrm{SU}(2)\otimes \mathrm{SU}(2)$, generated by $\{H_j\}_{j=1}^4$,
can be implemented very fast by hard pulses that excite each of the spins individually.

Generally, any $U\in \mathrm{SU}(4)$ has the Cartan decomposition \cite{H1978},
\be\label{Cartan}
U=K_1\exp[\frac{i}{2}(a_1\sigma_x\otimes\sigma_x+a_2\sigma_y\otimes\sigma_y+a_3\sigma_z\otimes\sigma_z)]K_2,
\ee
where $K_j\in\mathrm{SU}(2)\ot\mathrm{SU}(2)$, $j=1,2$, and the real $a_k$, $k=1,2,3$, are called the Cartan coordinates of $U$.
In \cite{Navin2001}, the minimum time $t^*$ required to implement a gate $U$ is shown to be the smallest possible value of $\frac{1}{\pi J}\sum_{k=1}^3|a_k|$, i.e.,
\be\label{mini-time}t^*=\frac{1}{\pi J}\min\sum_{k=1}^3|a_k|.\ee
The Cartan coordinates are not unique. They vary with the choices of $K_j\in\mathrm{SU}(2)\ot\mathrm{SU}(2)$, $j=1,2$.
Hence the minimum time to implement the quantum gate requires one to find all the possible Cartan coordinates.

In this paper, we propose an improved approach to deal with the minimum time problem of implementing an arbitrary two-qubit gate in two heteronuclear spins systems.
We introduce more local invariants and classify the quantum gates into four classes. We derive the analytical formula of the
minimum time to implement the quantum gates in each class.
Our strategy has two steps. We first show that there are at most four possible classes of two-qubit gates,
once the two invariants defined in \cite{ZhJun2003} are fixed. Then by simply evaluating our new invariants, the class of an arbitrary
quantum gate belonging to is identified, and the minimum time to implement the gate is obtained, thus solving completely the optimal control problem.

\section{Classification of unitary operators $U\in \mathrm{SU}(4)$}

Two unitary transformations $U, U'\in \mathrm{SU}(4)$ on the space $\mathbb{C}^2\ot \mathbb{C}^2$
are called locally equivalent if they differ only by local operations, i.e., there exist local gates $K_1, K_2\in \mathrm{SU}(2)\otimes \mathrm{SU}(2)$ such that
$U'=K_1UK_2$. Denote
\begin{equation}\label{e:rep-c}
[a_1,a_2,a_3]=\exp\left\{\frac{i}{2}(a_1\sigma_x\otimes\sigma_x
+a_2\sigma_y\otimes\sigma_y+a_3\sigma_z\otimes\sigma_z)\right\}.
\end{equation}
Then the Cartan decomposition (\ref{Cartan}) can be written as $U=K_1[a_1, a_2, a_3]K_2$, where $a_k=a_k(U)$, $k=1,2,3$.
Clearly the Cartan coordinates $a_k(U)$, $k=1,2,3$, are multi-valued functions of $U$.
To determine $a_k$, $k=1,2,3$, consider
the Bell basis: $|\Phi^+\ra=1/\sqrt{2}(|00\ra+|11\ra)$, $|\Phi^-\ra=i/\sqrt{2}(|01\ra+|10\ra)$,
$|\Psi^+\ra=1/\sqrt{2}(|01\ra-|10\ra)$, and $|\Psi^-\ra=i/\sqrt{2}(|00\ra-|11\ra)$.
The transition matrix from the standard computational basis $\{|00\ra,|01\ra,|10\ra,|11\ra\}$ to the Bell basis $\{|\Phi^+\ra, |\Phi^-\ra,
|\Psi^+\ra, |\Psi^-\ra\}$ is given by the following well-known unitary matrix
$$
Q=\frac{1}{\sqrt{2}}\left(\ba{cccc}1&0&0&i\\ 0&i&1&0\\ 0&i&-1&0\\ 1&0&0&-i\ea\right).
$$

With respect to the Bell basis, any two-qubit gate $U$ performs as the matrix $Q^\dag UQ$. We call
$B(U):=Q^\dag UQ$ the Bell form of $U$. For two-qubit local gate $K\in\mathrm{SU}(2)\otimes \mathrm{SU}(2)$, one always has that $B(K)=Q^\dag KQ\in \mathrm{SO}(4)$.
Therefore two unitary matrices $U$ and $U'$ are locally equivalent if and only if $B(U)=Q^{\dagger}UQ$ and $B(U')=Q^{\dagger}U'Q$ are
orthogonally equivalent \cite{J2015}, i.e., $B(U')=O_1B(U)O_2$ for some special orthogonal matrices $O_1, O_2\in \mathrm{SO}(4)$.

For any two-qubit gate $U\in \mathrm{SU}(4)$, we have
\be\label{Bell} B(U)=Q^\dag UQ=Q^\dag K_1[a_1,a_2,a_3]K_2Q=O_1Q^\dag [a_1,a_2,a_3]QO_2, \ee
where $O_j=B(K_j)=Q^\dag K_jQ\in \mathrm{SO}(4)$, $j=1,2$. In other words,
$B(U)$ is orthogonally equivalent to $Q^\dag [a_1,a_2,a_3]Q$.
Moreover, the Bell matrix form of $[a_1,a_2,a_3]$
is diagonal:
\begin{eqnarray}
B([a_1,a_2,a_3])&=&Q^\dag [a_1,a_2,a_3]Q=\exp\left\{\frac{i}{2}(a_1\sigma_z\otimes I
+a_2I\otimes\sigma_z+a_3\sigma_z\otimes\sigma_z)\right\}\nonumber\\
&=&\mathrm{diag}\left(e^{ib_1}, e^{ib_2},
e^{ib_3}, e^{ib_4}\right),
\label{NFB}\end{eqnarray}
where
\be\label{b-com}
b_1=\frac{a_1-a_2+a_3}{2},\ \ b_2=\frac{a_1+a_2-a_3}{2},\ \ b_3=-\frac{a_1+a_2+a_3}{2},
\ \ b_4=\frac{-a_1+a_2+a_3}{2}.\ee
Let
\be\label{Square}
m(U)=B(U)^TB(U)=O_2^TB([a_1,a_2,a_3])^2O_2.\ee

The following quantities are local invariants such that any locally equivalent two-qubit gates should have the same value \cite{ZhJun2003}:
\begin{eqnarray}\label{inv}
G_1(U)=\frac{\mathrm{Tr}^2(m(U))}{16}\equiv a+ib, ~~~
G_2(U)=(\mathrm{Tr}^2(m(U))-\mathrm{Tr}(m^2(U)))/4\equiv c,
\end{eqnarray}
where $a=\cos^2 a_1\cos^2 a_2\cos^2 a_3-\sin^2 a_1\sin^2 a_2\sin^2 a_3$, $b=\frac{1}{4}\sin 2a_1\sin 2a_2\sin 2a_3$ and
$c=4\cos^2 a_1\cos^2 a_2\cos^2 a_3-4\sin^2 a_1\sin^2 a_2\sin^2 a_3-\cos 2a_1\cos 2a_2\cos 2a_3$.
To find the solution $a_1$, $a_2$ and $a_3$ in terms of the local invariants $a$, $b$ and $c$,
the following cubic equation is concerned \cite{Yu2013},
\begin{equation}\label{cubic}
x^3+px^2+qx+r=(x-\sin^2 a_1)(x-\sin^2 a_2)(x-\sin^2 a_3)=0,
\end{equation}
where
\begin{equation}\label{pqr}
\ba{l}\displaystyle
p =-(1+\frac{1-c}{2}),~~~
q=\sqrt{a^2+b^2}+\frac{1-c}{2},~~~
\displaystyle
r=-\frac{1}{2}(\sqrt{a^2+b^2}-a).
\ea
\end{equation}
The three solutions of (\ref{cubic}) are given by $c_k=\sin^2 a_k$, $k=1,2,3$, which give the relations between $\{a_k\}$ and the invariants $a$, $b$ and $c$.

Denote
\be\label{alpha}
\alpha_k=\arcsin{|\sin a_k|}\in [0,\frac{\pi}{2}], \ k=1,2,3.
\ee
Note that the representative $[a_i,a_j,a_k]$ is locally equivalent to $[a_1,a_2,a_3]$ under any permutation $(i,j,k)$ of $(1,2,3)$ \cite{ZhJun2003}.
We can always assume that $\frac{\pi}{2}\geq\alpha_1\geq\alpha_2\geq\alpha_3\geq0$.
Then any $a_k$ is seen to take the following possible values:
\be\label{C--V--8}
2n\pi+\alpha_k, ~~2n\pi+\pi+\alpha_k,~~ 2n\pi+\pi-\alpha_k, ~~2n\pi-\alpha_k,~~ k=1,2,3,
\ee
where $n$ is an arbitrary integer. $B([a_1,a_2,a_3])$ is periodic with a period $4\pi$ for each $a_k$. To find the minimal value of $\sum_{k=1}^3|a_k|$, the values of $a_k$ can be confined in $[-2\pi, 2\pi]$.
Hence, every $a_k$ can have 8 possible values $\pm\alpha_k$, $\pm(\pi+\alpha_k)$, $\pm(\pi-\alpha_k)$ and $\pm(2\pi-\alpha_k) $. Therefore, for fixed $G_1$ and $G_2$, the triple $(a_1,a_2,a_3)$ has $8^3=512$ choices.
Nevertheless, since
\begin{equation}
B([a_1, a_2, a_3])=-B([a_1+2\pi, a_2, a_3])=-B([a_1+\pi, a_2+\pi, a_3]),
\end{equation}
$[a_1, a_2, a_3]$, $[a_1+2\pi, a_2, a_3]$ and $[a_1+\pi, a_2+2\pi, a_3]$ are locally equivalent.
Furthermore, it follows from the symmetry of $a_k$'s that all $[a_1, a_2+2\pi, a_3]$, $[a_1, a_2, a_3+2\pi]$, $[a_1, a_2+\pi, a_3+\pi]$ and $[a_1+\pi, a_2, a_3+\pi]$ are locally equivalent, thus
cutting down the possible choices of $(a_1, a_2, a_3)$ to $4^3=64$.
Noting that
\begin{align*}
B([-a_1, -a_2, a_3])&=\mathrm{diag}(e^{ib_4}, e^{ib_3},
	e^{ib_2}, e^{ib_1})\\
	&={\cal J}B([a_1, a_2, a_3]),
\end{align*}
where ${\cal J}\in \mathrm{SO}(4)$ is the skew diagonal matrix, we have that $[a_1, a_2, a_3]$ is locally equivalent to $[-a_1, -a_2, a_3]$. Therefore,
the classes $[a_1,a_2,a_3]$, $[\pi-a_1,\pi-a_2,a_3]$, $[-a_1,-a_2,a_3]$, $[\pi-a_1,a_2,\pi-a_3]$, $[-a_1,a_2,-a_3]$, $[a_1,\pi-a_2,\pi-a_3]$ and $[a_1,-a_2,-a_3]$ are all locally equivalent.
At last, for fixed $G_1$ and $G_2$, with $\alpha_k=\arcsin|\sin a_k|\in [0, \pi/2]$, $k=1,2,3$, $[a_1, a_2, a_3]$ can only be one of the four local classes:
\begin{equation}\label{e:4class}
[\alpha_1,\alpha_2,\alpha_3], ~~ [-\alpha_1,\alpha_2,\alpha_3],~~[\pi-\alpha_1,\alpha_2,\alpha_3],~~[-\pi+\alpha_1,\alpha_2,\alpha_3].
\end{equation}

We list the 64 choices in tables \ref{TAB1} and table \ref{TAB2}.
The quantity $\sum_{k=1}^3|a_k|$ is the same for two-qubit gates in classes I and II (III and IV).
We have set in tables \ref{TAB1} and table \ref{TAB2},
\be\label{beta}
\beta_1=\frac{\alpha_1-\alpha_2+\alpha_3}{2},\ \beta_2=\frac{\alpha_1+\alpha_2-\alpha_3}{2},\
\beta_3=-\frac{\alpha_1+\alpha_2+\alpha_3}{2},\ \beta_4=\frac{-\alpha_1+\alpha_2+\alpha_3}{2}.
\ee

\begin{table}[H]
  \centering
  \caption{Classification of 2-qubit gates for given $G_1$ and $G_2$}
\label{TAB1}
  \begin{tabular}{||c|c|c||c|c|c||c|c||}
   \hline
   \multicolumn{3}{||c||}{Class\quad I}&\multicolumn{3}{c||}{Class\quad II}&\multicolumn{2}{c||}{}\\
   \hline
   $a_1$&$a_2$&$a_3$&$a_1$&$a_2$&$a_3$&$\sum_{k=1}^3|a_k|$& $t^*$\\
   \hline
   $\alpha_1$&$\alpha_2$&$\alpha_3$&$-\alpha_1$&$\alpha_2$&$\alpha_3$
   &\multirow{4}*{$-2\beta_3$}&\multirow{16}*{$-\frac{2\beta_3}{\pi J}$}\\
   \cline{1-6}
   $-\alpha_1$&$-\alpha_2$&$\alpha_3$&$\alpha_1$&$-\alpha_2$&$\alpha_3$&&\\
   \cline{1-6}
   $-\alpha_1$&$\alpha_2$&$-\alpha_3$&$\alpha_1$&$\alpha_2$&$-\alpha_3$&&\\
   \cline{1-6}
   $\alpha_1$&$-\alpha_2$&$-\alpha_3$&$-\alpha_1$&$-\alpha_2$&$-\alpha_3$&&\\
   \cline{1-7}
   $\pi-\alpha_1$&$\pi-\alpha_2$&$\alpha_3$&$-\pi+\alpha_1$&$\pi-\alpha_2$&$\alpha_3$
   &\multirow{4}*{$2\pi-2\beta_2$}&\\
   \cline{1-6}
   $-\pi+\alpha_1$&$-\pi+\alpha_2$&$\alpha_3$&$\pi-\alpha_1$&$-\pi+\alpha_2$&$\alpha_3$&&\\
   \cline{1-6}
   $-\pi+\alpha_1$&$\pi-\alpha_2$&$-\alpha_3$&$\pi-\alpha_1$&$\pi-\alpha_2$&$-\alpha_3$&&\\
   \cline{1-6}
   $\pi-\alpha_1$&$-\pi+\alpha_2$&$-\alpha_3$&$-\pi+\alpha_1$&$-\pi+\alpha_2$&$-\alpha_3$&&\\
   \cline{1-7}
   $\pi-\alpha_1$&$\alpha_2$&$\pi-\alpha_3$&$-\pi+\alpha_1$&$\alpha_2$&$\pi-\alpha_3$
   &\multirow{4}*{$2\pi-2\beta_1 $}&\\
   \cline{1-6}
   $-\pi+\alpha_1$&$-\alpha_2$&$\pi-\alpha_3$&$\pi-\alpha_1$&$-\alpha_2$&$\pi-\alpha_3$&&\\
   \cline{1-6}
   $-\pi+\alpha_1$&$\alpha_2$&$-\pi+\alpha_3$&$\pi-\alpha_1$&$\alpha_2$&$-\pi+\alpha_3$&&\\
   \cline{1-6}
   $\pi-\alpha_1$&$-\alpha_2$&$-\pi+\alpha_3$&$-\pi+\alpha_1$&$-\alpha_2$&$-\pi+\alpha_3$&&\\
   \cline{1-7}
   $\alpha_1$&$\pi-\alpha_2$&$\pi-\alpha_3$&$-\alpha_1$&$\pi-\alpha_2$&$\pi-\alpha_3$
   &\multirow{4}*{$2\pi-2\beta_4$}&\\
   \cline{1-6}
   $-\alpha_1$&$-\pi+\alpha_2$&$\pi-\alpha_3$&$\alpha_1$&$-\pi+\alpha_2$&$\pi-\alpha_3$&&\\
   \cline{1-6}
   $-\alpha_1$&$\pi-\alpha_2$&$-\pi+\alpha_3$&$\alpha_1$&$\pi-\alpha_2$&$-\pi+\alpha_3$&&\\
   \cline{1-6}
   $\alpha_1$&$-\pi+\alpha_2$&$-\pi+\alpha_3$&$-\alpha_1$&$-\pi+\alpha_2$&$-\pi+\alpha_3$&& \\
   \hline
\end{tabular}
 \end{table}

\begin{table}[H]
  \centering
  \caption{Classification of 2-qubit gates for given $G_1$ and $G_2$}
\label{TAB2}
  \begin{tabular}{||c|c|c||c|c|c||c|c||}
   \hline
   \multicolumn{3}{||c||}{Class\quad III}&\multicolumn{3}{c||}{Class\quad IV}&\multicolumn{2}{c||}{}\\
   \hline
   $a_1$&$a_2$&$a_3$&$a_1$&$a_2$&$a_3$&$\sum_{k=1}^3|a_k|$& $t^*$\\
   \hline
   $\pi-\alpha_1$&$\alpha_2$&$\alpha_3$
   &$-\pi+\alpha_1$&$\alpha_2$&$\alpha_3$&\multirow{4}*{$\pi+2\beta_4$}&\multirow{16}*{$\frac{\pi+2\beta_4}{\pi J}$}\\ \cline{1-6}
   $-\pi+\alpha_1$&$-\alpha_2$&$\alpha_3$
   &$\pi-\alpha_1$&$-\alpha_2$&$\alpha_3$&&\\  \cline{1-6}
   $-\pi+\alpha_1$&$\alpha_2$&$-\alpha_3$
   &$\pi-\alpha_1$&$\alpha_2$&$-\alpha_3$&&\\  \cline{1-6}
   $\pi-\alpha_1$&$-\alpha_2$&$-\alpha_3$
   &$-\pi+\alpha_1$&$-\alpha_2$&$-\alpha_3$&&\\  \cline{1-7}
    $\alpha_1$&$\pi-\alpha_2$&$\alpha_3$
   &$-\alpha_1$&$\pi-\alpha_2$&$\alpha_3$&\multirow{4}*{$\pi+2\beta_1$}&\\ \cline{1-6}
   $-\alpha_1$&$-\pi+\alpha_2$&$\alpha_3$
   &$\alpha_1$&$-\pi+\alpha_2$&$\alpha_3$&&\\  \cline{1-6}
   $-\alpha_1$&$\pi-\alpha_2$&$-\alpha_3$
   &$\alpha_1$&$\pi-\alpha_2$&$-\alpha_3$&&\\  \cline{1-6}
   $\alpha_1$&$-\pi+\alpha_2$&$-\alpha_3$
   &$-\alpha_1$&$-\pi+\alpha_2$&$-\alpha_3$&&\\   \cline{1-7}
   $\alpha_1$&$\alpha_2$&$\pi-\alpha_3$
   &$-\alpha_1$&$\alpha_2$&$\pi-\alpha_3$&\multirow{4}*{$\pi+2\beta_2$}&\\ \cline{1-6}
   $-\alpha_1$&$-\alpha_2$&$\pi-\alpha_3$
   &$\alpha_1$&$-\alpha_2$&$\pi-\alpha_3$&&\\ \cline{1-6}
   $-\alpha_1$&$\alpha_2$&$-\pi+\alpha_3$
   &$\alpha_1$&$\alpha_2$&$-\pi+\alpha_3$&&\\ \cline{1-6}
   $\alpha_1$&$-\alpha_2$&$-\pi+\alpha_3$
   &$-\alpha_1$&$-\alpha_2$&$-\pi+\alpha_3$&&\\ \cline{1-7}
   $\pi-\alpha_1$&$\pi-\alpha_2$&$\pi-\alpha_3$
   &$-\pi+\alpha_1$&$\pi-\alpha_2$&$\pi-\alpha_3$&\multirow{4}*{$3\pi+2\beta_3$}&\\ \cline{1-6}
   $-\pi+\alpha_1$&$-\pi+\alpha_2$&$\pi-\alpha_3$
   &$\pi-\alpha_1$&$-\pi+\alpha_2$&$\pi-\alpha_3$&&\\ \cline{1-6}
   $-\pi+\alpha_1$&$\pi-\alpha_2$&$-\pi+\alpha_3$
   &$\pi-\alpha_1$&$\pi-\alpha_2$&$-\pi+\alpha_3$&&\\ \cline{1-6}
   $\pi-\alpha_1$&$-\pi+\alpha_2$&$-\pi+\alpha_3$
   &$-\pi+\alpha_1$&$-\pi+\alpha_2$&$-\pi+\alpha_3$&&\\ \hline
 \end{tabular}
 \end{table}

We have shown that, for fixed invariants $G_1$ and $G_2$, any two-qubit gates belong to
one of the classes I-IV listed in Tables 1 and 2. The
minimum time for implementing two-qubit gates belonging to classes I and II (or III and IV) is the same.
Next we need to distinguish the two-qubit gates belonging to classes I and II from that belonging to classes III and IV.

Decompose $B(U)$ from \eqref{Bell} into its real and imaginary parts:
\begin{eqnarray}\nonumber B(U)&=&O_1Q^\dag [a_1,a_2,a_3]QO_2=B_1(U)+iB_2(U)\\
\label{RE,IM}&=&O_1B_1([a_1,a_2,a_3])O_2+iO_1B_2([a_1,a_2,a_3])O_2,\end{eqnarray}
where $B_1(\cdot)$ and $B_2(\cdot)$ are the real and imaginary parts of $B(\cdot)$, respectively. We define
\begin{eqnarray}\label{Re-det}
G_3(U)&=&\mathrm{det}B_1(U)=\mathrm{det}\left(\frac{Q^\dag UQ+Q^T\overline{U}\overline{Q}}{2}\right)
=\prod_{k=1}^4\cos b_k,  \\[2mm]\label{Im-det}
G_4(U)&=&\mathrm{Tr}\left(B_1(U)B_2^T(U)\right)=\frac{1}{2}\sum_{k=1}^4\sin 2b_k,
\end{eqnarray}
where $b_j,\ j=1,2,3,4$ are defined in \eqref{b-com}.
One can verify that if $B(U')=O_1B(U)O_2$, then $B_k(U')=O_1B_k(U)O_2,\ k=1,2$ and
$B_1(U')B_2^T(U')=O_1B_1(U')B_2^T(U)O_1^T$ for $O_1, O_2\in\mathrm{SO}(4)$.
Hence, the quantities $G_3$ and $G_4$ are indeed local invariants.

It is direct to compute that
$$ \begin{array}{rlrl}
G_3([\alpha_1,\alpha_2,\alpha_3])&=\prod_{k=1}^4\cos \beta_k,
  \ \ \ \ &G_4([\alpha_1,\alpha_2,\alpha_3])&=\frac12\sum_{k=1}^4\sin 2\beta_k.\\
   G_3([-\alpha_1,\alpha_2,\alpha_3])&=\prod_{k=1}^4\cos \beta_k,
    \ \ \ \ &G_4([-\alpha_1,\alpha_2,\alpha_3])&=-\frac12\sum_{k=1}^4\sin 2\beta_k.\\
   G_3([\pi-\alpha_1,\alpha_2,\alpha_3])&=\prod_{k=1}^4\sin \beta_k,
   \ \ \ \  &G_4([\pi-\alpha_1,\alpha_2,\alpha_3])&=\frac12\sum_{k=1}^4\sin 2\beta_k.\\
   G_3([-\pi+\alpha_1,\alpha_2,\alpha_3])&=\prod_{k=1}^4\sin \beta_k,
\ \ \ \ &G_4([-\pi+\alpha_1,\alpha_2,\alpha_3])&=-\frac12\sum_{k=1}^4\sin 2\beta_k.
\end{array}$$
Therefore, we conclude that if $G_3=\prod_{k=1}^4\cos \beta_k$ ($G_3=\prod_{k=1}^4\sin \beta_k$), then
the corresponding two-qubit gate belongs to class I and II (III and IV).
Moreover, if $G_4=\frac12\sum_{k=1}^4\sin 2\beta_k$ ($G_4=-\frac12\sum_{k=1}^4\sin 2\beta_k$),
then the two-qubit gate belongs to class I and III (II and IV).
Altogether, the invariants $G_i$, $i=1,...,4$, can identify which class a two-qubit gate belongs to, see Table 3.

\begin{table}[H]
\centering
\caption{Classes given by the values of $G_3$ and $G_4$}
\label{TAB3}
    \begin{tabular}{||c||c|c|c||}
   \hline
   class &$G_3$&$G_4$\\
   \hline
   I&$\prod_{k=1}^4\cos\beta_k$
   &$\frac{1}{2}(\sin2\beta_1+\sin2\beta_2+\sin2\beta_3+\sin2\beta_4)$\\
   \hline
   II&$\prod_{k=1}^4\cos\beta_k$
   &$-\frac{1}{2}(\sin2\beta_1+\sin2\beta_2+\sin2\beta_3+\sin2\beta_4)$\\
   \hline
   III&$\prod_{k=1}^4\sin\beta_k$
   &$\frac{1}{2}(\sin2\beta_1+\sin2\beta_2+\sin2\beta_3+\sin2\beta_4)$\\
   \hline
   IV&$\prod_{k=1}^4\sin\beta_k$
   &$-\frac{1}{2}(\sin2\beta_1+\sin2\beta_2+\sin2\beta_3+\sin2\beta_4)$\\
   \hline
 \end{tabular}
 \end{table}

\section{The minimum time $t^*$ to implement a two-qubit gate $U$}

We now present our method to compute the optimal time to implement a two-qubit gate $U$.
For given $U$, one first computes $G_1$ and $G_2$ and hence $a$, $b$ and $c$ by (\ref{inv}).
Solving (\ref{cubic}) one gets three solutions $c_1\geq c_2\geq c_3$ in terms of $a$, $b$ and $c$.
From \eqref{alpha}, we have $\alpha_k=\arcsin\sqrt{c_k}$, $k=1,2,3$, and then $\beta_k$, $k=1,2,3,4$, by \eqref{beta}.
Next, one computes $G_3(U)$ using  \eqref{Re-det}.  If $G_3(U)=\prod_{k=1}^4\cos\beta_k$, then $U$ belongs to the Class I or Class II.
If $G_3(U)=\prod_{k=1}^4\sin\beta_k$, then $U$ belongs to Class III or Class IV.

Tables \ref{TAB1} and \ref{TAB2} show that the minimum value of $\sum_{k=1}^3|a_k|$
is $\min\{2\pi-2\beta_1, 2\pi-2\beta_2, -2\beta_3,2\pi-2\beta_4\}$
for gates in classes I and II, and $\min\{\pi+2\beta_1, \pi+2\beta_2, 3\pi+2\beta_3,\pi+2\beta_4\}$ for gates in classes III and IV.
Since $0\leq\alpha_3\leq\alpha_2\leq\alpha_1\leq\frac{\pi}{2}$,
the minimum values of $\sum_{k=1}^3|a_k|$ for classes I, II and classes III, IV are $-2\beta_3$ and $\pi+2\beta_4$, respectively.
Therefore, if $U$ belongs to Class I or Class II, the minimum time for implementing $U$ is $t^*=-\frac{2\beta_3}{\pi J}$.
If $U$ belongs to Class III or Class IV, the minimum time for implementing $U$ is $t^*=\frac{\pi+2\beta_4}{\pi J}$.

The role of global phase in quantum evolution operators has been studied from
various aspects. For example, the effect resulting from the such phase difference is the
overall phase change acquired after the $2\pi$ rotation of a
particle \cite{6,7}, which distinguishes fermions from bosons \cite{3}, as observed in experimentally via interferometric approaches \cite{8,9,10}.
Recently, the distinctions among operations differ by a global phase have been studied \cite{11,12,13,14,15}.
In \cite{11}, the relations between the global phase of a SU(2) operation and the corresponding optimal time to
realize such an operation has been derived.
Before some detailed examples, we first present below a systematic analysis on the effect of global phase on the optimal time
for SU(4) operators.

For the case of $U\in \mathrm{SU(4)}$, the global phase can only be $i=\sqrt{-1}$ due to that the determinant $\det(U)=1$.
Assume that $U\in \mathrm{SU(4)}$ has Cartan decomposition $U=K_1[a_1,a_2,a_3]K_2$,
where $K_1,K_2\in \mathrm{SU(2)}$, $|a_1|\geq a_2\geq a_3\geq0$. Since
$iI_4=\exp[i(\pm\pi/2)\sigma_\gamma\otimes\sigma_\gamma](\pm i\sigma_\gamma)\otimes (-i\sigma_\gamma)$, where $\gamma=x,y,z$,
the Cartan decomposition of $iU$ has the form,
\begin{eqnarray}
iU\nonumber&=&K_1\exp[\frac{i}{2}(a_1\sigma_x\otimes\sigma_x+a_2\sigma_y\otimes\sigma_y
+a_3\sigma_z\otimes\sigma_z)](iI_4)K_2\\
\nonumber&=&K_1\exp[\frac{i}{2}(a_1\sigma_x\otimes\sigma_x+a_2\sigma_y\otimes\sigma_y
+a_3\sigma_z\otimes\sigma_z)]\exp[\frac{i\pi}{2}\sigma_\gamma\otimes\sigma_\gamma] \widetilde{K}_\gamma\\
\label{CD-iU}&=&K_1\exp[\frac{i}{2}(a_1\sigma_x\otimes\sigma_x+a_2\sigma_y\otimes\sigma_y
+a_3\sigma_z\otimes\sigma_z\pm\pi\sigma_\gamma\otimes\sigma_\gamma)]\widetilde{K}_\gamma,
\end{eqnarray}
where $\widetilde{K}_\gamma=(\pm i\sigma_\gamma)\otimes (-i\sigma_\gamma)K_2$, $\gamma=x,y,z.$

Recall that $G_3(U)=\det(B_1(U))$ and $G_4(U)=\mathrm{Tr}(B_1(U)B_2^T(U))$, where $B_1(U)$ and $B_2(U)$ are given by $B(U)=B_1(U)+iB_2(U)$.
We have $B(iU)=-B_2(U)+iB_1(U)$, $G_3(iU)=\det(B_2(U))$ and $G_4(iU)=-\mathrm{Tr}(B_2(U)B_1^T(U))=-G_4(U)$.
Now if $U$ is in Class I or Class II in Table \ref{TAB1}, then $iU$ is in Class III or IV in Table \ref{TAB2}, respectively, and vice versa. Therefore, if $a_1\geq0$,
$K_1\exp[\frac{i}{2}((a_1-\pi)\sigma_x\otimes\sigma_x+a_2\sigma_y\otimes\sigma_y
+a_3\sigma_z\otimes\sigma_z]\widetilde{K}_x$ is the optimal decomposition of $iU$. If $a_1<0$, $K_1\exp[\frac{i}{2}((a_1+\pi)\sigma_x\otimes\sigma_x+a_2\sigma_y\otimes\sigma_y
+a_3\sigma_z\otimes\sigma_z]\widetilde{K}_x$ is the optimal decomposition of $iU$.
In particular, if $\frac{\pi}{2}=|a_1|\geq a_2\geq a_3\geq0$, $t^*(U)=t^*(iU)$.

Moreover, from tables \ref{TAB1} and \ref{TAB2}, $[a_1,a_2,a_3]$ is in Class I (resp. III) if and only if $[-a_1,-a_2,-a_3]$ is in Class II (resp. IV).
Simple computation gives that $G_1(U^\dag)=\overline{G_1(U)}$, $G_2(U^\dag)=G_2(U)$, $G_3(U^\dag)=G_3(U)$ and $G_4(U^\dag)=-G_4(U)$.
As the values of $\sin^2a_k$, $k=1,2,3$, only depend on $G_2(U)$ and the real part and the modulus of $G_1(U)$  from \eqref{pqr}, $U$ and $U^\dag$ give rise to the same values of $\sin^2 a_k$, $k=1,2,3$.
Hence, if $U$ is in Class I, then $U^\dag$ is in Class II, and $iU^\dag$ is in Class III, $iU$ is in Class IV, implying that
$U$ and $U^\dag$ have the same optimal time.

We present next some detailed examples including the ones considered in the literature \cite{ZhJun2003,Yu2013} to show the roles played by global phase in optimal time.

\textbf{Example 1. $I_4$ vs. $iI_4$.}

We have $G_1(U)=1$ and $G_2(U)=0$ for both $U=I_4$ and $U=iI_4$ from \eqref{Bell}, \eqref{Square} and \eqref{inv}. Then $\sin^2a_i=0$,  $i=1,2,3$ from \eqref{cubic} and \eqref{pqr}.
Now from \eqref{alpha} we have $(\alpha_1, \alpha_2, \alpha_3)=(0, 0, 0)$. Then we have $(\beta_1, \beta_2, \beta_3, \beta_4)=(0, 0, 0, 0)$ from \eqref{beta}, and
$\prod_{k=1}^4\cos\beta_k=1$, $\prod_{k=1}^4\sin\beta_k=0$ and $\sum_{k=1}^4\sin 2\beta_k=0$.
Since $G_3(I)=1=\prod_{k=1}^4\cos\beta_k$  and $G_3(iI)=0=\prod_{k=1}^4\sin\beta_k$ from \eqref{Re-det}, $I_4$ belongs to classes I and II, and $iI_4$ belongs to classes III and IV from Table \ref{TAB3}.
From Table \ref{TAB1}, Table \ref{TAB2} and \eqref{mini-time},
we have the minimum time $t^*$ required to implement $I_4$ and $iI_4$ are zero and $\frac{1}{J}$, respectively. 

{\it Remark} From the result of \cite{Yu2013}, the minimum time required to implement $I_4$ and $iI_4$ are both zero.

\textbf{Example 2. Controlled-NOT gate $U_{C_{NOT}}$}

The gate $U_{C_{NOT}}$ is given by $U_{C_{NOT}}=e^{\frac{i\pi}4}(|0\ra\la0|\otimes I+|1\ra\la1|\otimes\sigma_x)$.
For $U_{C_{NOT}}$ one has \cite{Yu2013} $\sin^2 a_1=1$, $\sin^2a_2=\sin^2a_3=0$. Hence
$(\alpha_1, \alpha_2, \alpha_3)=(\frac{\pi}{2}, 0, 0)$ and $(\beta_1, \beta_2, \beta_3, \beta_4)=(\frac{\pi}{4}, \frac{\pi}{4}, -\frac{\pi}{4}, -\frac{\pi}{4})$.
Here since $\pi-\alpha_1=\alpha_1$ and $\alpha_2=\alpha_3=0$, Class I and Class III have the same $[a_1, a_2, a_3]$ (the same is true for Class II and Class IV), see Table \ref{TAB1} and Table \ref{TAB2}.
Therefore, without computing $G_3$, $G_4$ and $\prod_{k=1}^4\cos\beta_k$, $\prod_{k=1}^4\sin\beta_k$, we can conclude that the minimal value of $\sum_{i=1}^3|a_i|$ is $\frac{\pi}{2}$
and the minimal time required to implement controlled-NOT gate is $\frac{1}{2J}$.

\textbf{Example 3. $U_{SWAP}$ vs. $iU_{SWAP}$.}

The SWAP gate $U_{SWAP}=\frac{1}{2}e^{i\frac{\pi}{4}}\left(\begin{array}{cc}I+\sigma_z&\sigma_x-i\sigma_y\\ \sigma_x+i\sigma_y&I-\sigma_z\end{array}\right)$.
We have $G_1(U)=-1$, $G_2(U)=-3$ for both $U=U_{SWAP}$ and $iU_{SWAP}$, and $\sin^2a_k=1$, $k=1,2,3$ \cite{Yu2013}. Then
$\alpha_1=\alpha_2=\alpha_3=\frac{\pi}{2}$, $\beta_1=\beta_2=\beta_4=\frac{\pi}{4}$ and $\beta_3=-\frac{3\pi}{4}$.
Clearly, $\prod_{k=1}^4\cos\beta_k=\prod_{k=1}^4\sin\beta_k=-\frac{1}{4}$. Hence $G_3(U)=-\frac{1}{4}$  for  both $U_{SWAP}$ and $iU_{SWAP}$.
From Table \ref{TAB1} and Table \ref{TAB2}, the minimal time required to implement $U_{SWAP}$ or $iU_{SWAP}$ are both $\frac{3}{2J}$.

\textbf{Example 4. $\sqrt{\mathbf{SWAP}}$ gate $U_{\sqrt{\mathbf{SWAP}}}$.}

Consider the $U_{\sqrt{\mathbf{SWAP}}}$ gate,
$$
U_{\sqrt{\mathbf{SWAP}}}=e^{i\frac{\pi}8}\left(\begin{array}{cccc}1&0&0&0\\ 0&\frac{1-i}{2}&\frac{1+i}{2}&0\\ 0&\frac{1+i}{2}&\frac{1-i}{2}&0\\ 0&0&0&1\end{array}\right).
$$
From \cite{Yu2013}, we have $\sin^2a_k=\frac{1}{2}$, $k=1,2,3$. Hence $\alpha_k=\frac{\pi}{4},\ k=1,2,3$, $\beta_1=\beta_2=\beta_4=\frac{\pi}{8}$ and  $\beta_3=-\frac{3\pi}{8}$. Using \eqref{Re-det}, we get $$G_3(U_{\sqrt{\mathbf{SWAP}}})=\cos^3\frac{\pi}{8}\sin\frac{\pi}{8}=\prod_{k=1}^4\cos\beta_k$$
 and
$$G_3(iU_{\sqrt{\mathbf{SWAP}}})=-\sin^3\frac{\pi}{8}\cos\frac{\pi}{8}=\prod_{k=1}^4\sin\beta_k.$$
From Table \ref{TAB3}, $U_{\sqrt{\mathbf{SWAP}}}$ belongs to classes I or II, and $iU_{\sqrt{\mathbf{SWAP}}}$ to classes III or IV. Hence we obtain that
$t^*(U_{\sqrt{\mathbf{SWAP}}})=\frac{3}{4J}$ and $t^*(iU_{\sqrt{\mathbf{SWAP}}})=\frac{5}{4J}$.


\section{Conclusion}

Optimal time implementation of a quantum gate is one of the important tasks in quantum computation. The algorithm presented in \cite{Yu2013}, using the two local invariants $G_1$ and $G_2$, is inconclusive: it does not provide a conclusive answer even for the simple 2-qubit gate such as $iI_4$.
To completely settle the optimal time problem, we have introduced two new local invariants $G_3$ and $G_4$ in terms of the Bell form of 2-qubit gates.
We have shown that $G_1$, $G_2$ and $G_3$ are sufficient to calculate the optimal time to implement an arbitrary 2-qubit gate, which
provides an effective and decisive method to resolve the quantum optimal control problem.
As applications, we have used some well-known unitary gates to showcase our method in determination of the minimum time for implementing
these gates. Moreover, the effect of global phases on the minimum time to implement a quantum gate has been extensively analyzed.
Our results present a complete characterization of the optimal time problem in implementing an arbitrary two-qubit gate in two heteronuclear spins systems.

Recently in \cite{Ji-PRA98-062108} the authors studied the time-optimal control of independent spin-$1/2$ systems under simultaneous control. The optimal control
has been experimentally implemented by using zero-field nuclear magnetic resonance (NMR).
It would be also interesting to demonstrate our theoretical results experimentally in detailed quantum systems like NMR.

\bigskip

\noindent{\bf Acknowledgments}\, \, This work is supported by the NSF of China under grant Nos. 11531004, 11675113, 11701320, Shandong provincial NSF of China grant No. ZR2016AM04, Simons Foundation grant No. 523868, Beijing Municipal Commission of Education under grant No. KZ201810028042, and Beijing Natural Science Foundation (Z190005).

\end{document}